\documentclass[preprint]{ptephy_v1}

\usepackage{amssymb}
\usepackage{graphicx}
\numberwithin{equation}{section}

\newcommand{\dl}{\delta}
\newcommand{\cb}{\bar{c}}
\newcommand{\lb}[1]{\label{#1}}
\newcommand{\cl}{\mathcal{L}}
\newcommand{\bb}{\bar{B}}
\newcommand{\dlb}{\bar{\delta}}
\newcommand{\vp}{\varphi}
\newcommand{\ptl}{\partial}
\newcommand{\ve}{\varepsilon}
\def\vec#1{\mbox{\boldmath $#1$}}

\begin{document}

\title{The Gribov Horizon and Ghost Interactions in Euclidean Gauge Theories
}

\author{\name{\fname{Hirohumi} \surname{Sawayanagi}}{1}}

\address{\affil{1}{National Institute of Technology, Kushiro College, Kushiro, 084-0916, Japan}
\email{sawa@kushiro-ct.ac.jp}} 

\begin{abstract}
The effect of the Gribov horizon in Euclidean $SU(2)$ gauge theory is 
studied.  Gauge fields on the Gribov horizon yield zero modes of 
ghosts and anti-ghosts.  We show these zero modes can produce additional ghost 
interactions, and the Landau gauge changes to a nonlinear gauge effectively.  
In the infrared limit, however, the Landau gauge is recovered, and ghost zero modes 
may appear again.  We show ghost condensation happens in the nonlinear gauge, and 
the zero mode repetition is avoided.  
\end{abstract}

\subjectindex{B05,B06}

\maketitle

\section{Introduction}

  A perturbative calculation in gauge theories requires gauge fixing.  
However, in non-Abelian gauge theories, there is a problem of gauge 
copies \cite{gri}.  Gribov showed that gauge-equivalent copies exist in the 
Landau gauge 
\begin{equation}
 \ptl_{\mu}A_{\mu} = 0. \lb{101}
\end{equation}
In the Coulomb gauge, it was shown that almost all gauge transformations are 
responsible for gauge fixing degeneracies \cite{masnak}.  If gauge copies are 
connected by an infinitesimal gauge transformation with a gauge parameter 
$\ve (x)$, (\ref{101}) gives $\ptl_{\mu}D_{\mu} \ve (x)=0$.  That is, 
the Faddeev-Popov (FP) operator $-\ptl_{\mu}D_{\mu}$ has zero eigenvalues.  
The boundary that the lowest eigenvalue of the FP operator equals zero is called 
the (first) Gribov horizon $\ptl \Omega$.  The region inside 
$\ptl \Omega$, 
where eigenvalues of $-\ptl_{\mu}D_{\mu}$ are positive, is called the Gribov 
region $\Omega$.  In general, gauge copies may exist outside of 
$\Omega$ \cite{gri} and on the horizon \cite{bal}.  

  There are some ideas to solve the problem.  One of them is to restrict a 
functional integral in the Gribov region $\Omega$ \cite{gri,zw}.  
(Strictly speaking, there may be some copies in $\Omega$.  Hence more 
restricted region in $\Omega$, that is called a fundamental modular region 
(FMR) $\Lambda$, 
is considered \cite{vb}.)  Another idea is to 
sum over all gauge copies \cite{fuj,hir}.  For a solvable gauge model, it 
was shown that 
correct results are obtained by collecting all gauge copies \cite{flpr,fuj2}.  

  The Gribov horizon yield some effects.  In the first approach, the 
horizon perturbs gluons into shadow particles \cite{zw,ms}.  Even if the region 
is restricted to the FMR $\Lambda$, there are points that the boundary of 
$\Lambda$ touches the horizon $\ptl \Omega$ \cite{vb}.  These points give the 
singularity of the operator $1/\ptl_{\mu}D_{\mu}$.  As a result, the color 
Coulomb potential is enhanced and the confinement might be shown \cite{zw2}.  
In the second approach, gauge configurations on the Gribov 
horizon contribute in general, and the FP operator has zero modes.  These 
zero modes can cause a trouble in proving the gauge equivalence \cite{fuj3}.  
Thus physical effects of the horizon $\ptl \Omega$ are worth studying.  

  In this paper, we study the effect of these zero modes.  In the next 
section, we show that a pair of zero modes in the 
Landau gauge can yield additional ghost 
interactions.  If we require the BRS invariance, an effective Lagrangian 
becomes a Lagrangian in a nonlinear gauge.  In \S 3, the gauge $\ptl_{\mu}A_{\mu}\neq 0$ 
is considered.  If there is a pair of zero modes, the nonlinear gauge is 
realized as well.  We also show that the partition function does not vanish even if the FP operator 
yields a single zero mode.  In \S 4, the effect of a single zero mode  
is discussed in the Landau gauge.  In the low energy region, ghost condensation appears in the nonlinear gauge.  
The effect of the zero modes under the condensation is discussed in \S 5.  
\S 6 is devoted to summary.  In Appendix A, examples of zero modes 
in the Coulomb gauge are given in three dimensional space-time.  In Appendix B, the 
effective Lagrangian in \S 2 is derived by the use of a source term.  
The nonlinear gauge has two gauge parameters.  Renormalization group equations 
for these parameters are presented in Appendix C.  In Appendix D, symmetries 
in the nonlinear gauge are discussed.  

\section{Effect of ghost zero modes in the Landau gauge}

  We consider the $SU(2)$ gauge theory with structure constants $f^{ABC}$.  
Using the notations 
\[ F\cdot G=F^AG^A, \quad (F\times )^{AB}=f^{ACB}F^C, \quad
(F\times G)^A=f^{ABC}F^BG^C, \quad 
A=1,2,3, \]
a partition function in the Landau gauge is $Z_L=Z_{\alpha=0}$ with 
\begin{align}
  Z_{\alpha} = \int D\mu \, 
  e^{-\int dx (\cl_{\mathrm{inv}}+\cl_{\alpha})}, \quad 
  D\mu = DA_{\mu} DB Dc D\cb, \lb{201} \\
  \cl_{\mathrm{inv}} = \frac{1}{4}F_{\mu\nu}^2, \quad 
  \cl_{\alpha} = B\cdot \partial_{\mu}A_{\mu} 
  - \frac{\alpha}{2}B^2 + i\cb\cdot\ptl_{\mu}D_{\mu}c,  \lb{202}
\end{align}
where $i\cb\cdot\ptl_{\mu}D_{\mu}c=i\cb^A \ptl_{\mu}(\ptl_{\mu}+gA_{\mu}\times)^{AB}c^B$.  
The gauge condition (\ref{101}) leads to the relations 
\begin{equation}
 \ptl_{\mu}D_{\mu}=D_{\mu}\ptl_{\mu}, \quad 
\int dx\, i\cb\cdot\ptl_{\mu}D_{\mu}c=\int dx\, i(\ptl_{\mu}D_{\mu}\cb)\cdot c.
\lb{203}
\end{equation}
Namely, $\ptl_{\mu}D_{\mu}$ is hermitian, and its eigenvalues are 
real.  

   The eigenfunction $u_n$ with the eigenvalue $\lambda_n$ satisfies 
\begin{equation}
  -\ptl_{\mu}D_{\mu} u_n(x)=\lambda_n u_n(x).  \lb{204}
\end{equation}
When $A_{\mu}$ is on the first Gribov horizon, the lowest eigenvalue is $\lambda_0=0$ and 
$u_0(x)$ is a zero mode.  If we can make $u_0(x)$ complex, 
as (\ref{204}) leads to 
\begin{equation}
 -\ptl_{\mu}D_{\mu}u_n^*(x)=\lambda_n u_n^*(x),  \lb{205}
\end{equation}  
$u_0^*(x)$ is also a zero mode.  We assume a pair of zero modes $(u_0(x), u_0^*(x))$ 
exists.  Some examples of a zero-mode pair $(u_0(x), u_0^*(x))$ are presented in Appendix A.  
If $u_0$ is real, it may be a single zero mode.  An example of such a zero mode is given in 
Appendix A, and its effect is discussed in \S 4.  

  Now we expand the ghost $c$ as \footnote{We assume that eigenfunctions of the FP operator 
form an orthonormal complete set.  Strictly speaking, to ensure it, 
spaces and/or configurations of $A_{\mu}$ must be restricted.  We emphasize 
what is important here is that $c$ contains $\xi u_0, \xi^{\dagger}u_0^*$ and 
$\cb$ contains $\bar{\xi}u_0, \bar{\xi}^{\dagger}u_0^*$.}
\begin{equation}
 c(x) = \xi u_0(x) + \xi^{\dagger}u_0^*(x) + \cdots , 
 \lb{206}
\end{equation}
where $\xi$ and $\xi^{\dagger}$ are independent Grassmann variables.  Other modes, 
i.e. nonzero modes and a single zero mode, are not written explicitly.  In the same way, 
the property (\ref{203}) implies that the expansion 
\begin{equation}
 \cb(x) = \bar{\xi} u_0(x) + \bar{\xi}^{\dagger}u_0^*(x) + \cdots . 
 \lb{207}
\end{equation}
holds.   We note, if there are some pairs of 
zero modes $(u_0^j(x), u_0^{j*}(x)) \ (j=1,2,\cdots)$, 
$\xi u_0(x) + \xi^{\dagger}u_0^*(x)$ and 
$\bar{\xi} u_0(x) + \bar{\xi}^{\dagger}u_0^*(x)$ are replaced by 
$\sum_j [\xi_j u_0^j(x) + \xi_j^{\dagger}u_0^{j*}(x)]$ and 
$\sum_j [\bar{\xi_j} u_0^j(x) + \bar{\xi_j}^{\dagger}u_0^{j*}(x)]$, 
respectively.  However the discussion below is also applicable.  

  Eqs.(\ref{204}) and (\ref{205}) imply that the Lagrangian 
$\int dx\, i\cb\cdot\ptl_{\mu}D_{\mu}c$ does 
not contain the Grassmann variables 
$\xi, \xi^{\dagger}, \bar{\xi}$ and $\bar{\xi}^{\dagger}$.  However the 
measures $Dc$ and $D\cb$ contain $d\xi d\xi^{\dagger}$ and 
$d\bar{\xi} d\bar{\xi}^{\dagger}$, respectively.  
Since a Grassmann variable $\zeta$ satisfies 
\begin{equation}
  \int d\zeta\, \zeta^n = \left\{
\begin{array}{@{\,}ll}
1& (n=1) \\
0& (n=0,2,3,\cdots) 
\end{array}
\right. ,  \lb{208}
\end{equation}
the partition function vanishes: 
\[
  \int Dc D\cb \,e^{-\int dx \cl_{\alpha}} = 0.  
\]

    We know that fermions in an instanton background have zero modes.  These 
zero modes yield the additional interaction of fermions \cite{tH,tH2}.  
Likewise, 
the above ghost zero modes may produce additional ghost interactions, 
because 
\begin{equation}
  \int Dc D\cb\, \xi \xi^{\dagger} \bar{\xi} \bar{\xi}^{\dagger}\,
 e^{-\int dx \cl_{\alpha}} \neq 0.  \lb{209}
\end{equation}
From (\ref{206}) and (\ref{207}), we obtain 
\[ c^A c^B \cb^C \cb^D=\Psi^{ABCD}\xi \xi^{\dagger} \bar{\xi} 
\bar{\xi}^{\dagger} + \cdots,   \]
where $\Psi^{ABCD}=u_0^Au_0^Bu_0^{*C}u_0^{*D}$, and terms denoted by 
$\cdots$ lack some or all of $\xi, \xi^{\dagger}, \bar{\xi}$ and 
$\bar{\xi}^{\dagger}$.  Therefore (\ref{209}) leads to 
\begin{equation}
  \int Dc D\cb\, \sigma^{[AB][CD]}\Psi^{ABCD}
\xi \xi^{\dagger} \bar{\xi} \bar{\xi}^{\dagger}\, e^{-\int dx \cl_{\alpha}}=
\int Dc D\cb\,\sigma^{[AB][CD]}c^Ac^B\cb^C\cb^D\, e^{-\int dx \cl_{\alpha}},  \lb{210}
\end{equation}
where $\sigma^{[AB][CD]}$ is antisymmetric with respect to $A$ and $B$, and 
$C$ and $D$ as well.  Thus ghost zero modes produce effective ghost 
interactions.  

     Now we determine $\sigma^{[AB][CD]}$, and construct effective 
Lagrangians.  
The first candidate is $\sigma^{[AB][CD]}=f^{EAB}f^{ECD}
(=\delta^{AC}\delta^{BD}-\delta^{AD}\delta^{BC})$.  This choice gives the 
term 
\[  \sigma^{[AB][CD]}c^Ac^B\cb^C\cb^D=(\cb \times \cb)\cdot(c \times c) =
 -2(\cb \times c)\cdot (\cb \times c),   
\]
and (\ref{210}) becomes 
\begin{equation}
  \int Dc D\cb \, (\cb \times c)\cdot(\cb \times c)\, e^{-\int dx \cl_{\alpha}}.  
  \lb{211}
\end{equation}
From (\ref{208}), the equality 
\begin{equation}
  \int d\zeta\, e^{\zeta} = 1   \lb{212}
\end{equation}
holds.  Therefore, as in the instanton case \cite{cr}, (\ref{211}) is 
derived from the nonvanishing partition function 
\begin{equation}
 \int Dc D\cb\,e^{-\int dx \frac{K_1}{4}(i\cb \times c)^2}
 e^{-\int dx \cl_{\alpha}},  \lb{213}
\end{equation}
where $K_1$ is a dimensionless constant.  

     Interaction with other fields is also possible.  If we use 
$\sigma^{[AB][CD]}=B^EB^F(f^{EAC}f^{FBD}-f^{EBC}f^{FAD})$, 
\footnote{Instead of $B$, we can use $A_{\mu}$.  Examples are $F_{\mu\nu}$ 
and $\ptl_{\mu}A_{\mu}$.  However, using them, we cannot construct a Lagrangian 
which has mass dimension four (or lower than four) and has the off-shell BRS 
invariance.} we obtain the term 
\[ 
  \sigma^{[AB][CD]}c^Ac^B\cb^C\cb^D = 
-2[B\cdot(c\times \cb)][B\cdot(c\times \cb)], 
\]
and (\ref{210}) becomes 
\begin{equation}
  \int Dc D\cb \,[B\cdot(c\times \cb)][B\cdot(c\times \cb)]\, 
 e^{-\int dx \cl_{\alpha}}.  \lb{214}
\end{equation}
Taking account of (\ref{212}), we find (\ref{214}) is derived from 
\begin{equation}
  \int Dc D\cb\, e^{-\int dx K_2 B\cdot(\cb \times c)}e^{-\int dx \cl_{\alpha}}, 
 \lb{215}
\end{equation}
where $K_2$ is a dimensionless constant.  

     We can combine (\ref{213}) and (\ref{215}) in a BRS invariant form.  
Carrying out the BRS transformation 
\begin{equation}
  \dl_BA_{\mu}=D_{\mu}c, \quad \dl_Bc=-\frac{g}{2}c\times c, \quad 
 \dl_B\cb=iB,    \lb{216}
\end{equation}
we obtain 
\[
 \dl_B\left\{\frac{K_1}{2}(i\cb\times c)^2+K_2[B\cdot(\cb \times c)]\right\} 
 = (-iK_1-gK_2)(B\times c)\cdot(\cb\times c).  
\]
If we set $K_2=-\frac{i}{g}K_1=ig\alpha_2$, we get the BRS 
invariant effective Lagrangian  
\begin{equation}
  \cl_{\mathrm{eff}} = -\frac{\alpha_2}{2}(ig\cb \times c)^2
 + \alpha_2 B\cdot(ig\cb \times c)=
 \frac{\alpha_2}{2}B^2-\frac{\alpha_2}{2}\bb^2,  \lb{217}
\end{equation}
where $\bb = -B+ ig\cb \times c$, and $\alpha_2$ is a new dimensionless 
constant.  

     Here we used the property (\ref{208}) to derive the effective Lagrangian 
(\ref{217}).  In Appendix B, we derive it by using a source term.  

    Now we summarize the result.  In the Landau gauge, when the configuration 
$A_{\mu}$ on the Gribov horizon contribute to the partition function, the 
FP operator has zero modes.  If a pair of zero modes $(u_0(x), u_0^*(x))$ exists, the 
effective Lagrangian (\ref{217}) is produced.  From (\ref{201}) 
and (\ref{217}), we obtain the partition function 
\begin{align}
  Z &= Z_{\alpha=0}^{NL}  \notag  \\
  Z_{\alpha}^{NL} &=\int D\mu \, e^{-\int dx (\cl_{\mathrm{inv}} + \cl_{\alpha} + \cl_{\mathrm{eff}})} 
  =  \int D\mu \,e^{-\int dx (\cl_{\mathrm{inv}} + \cl_{NL})},  \lb{218}  \\
  \cl_{NL} &=  B\cdot \partial_{\mu}A_{\mu}+i\cb\cdot\partial_{\mu}D_{\mu}c-
 \frac{\alpha_1}{2}B^2-\frac{\alpha_2}{2}\bb^2,   \lb{219}
\end{align}
where $\alpha_1=\alpha-\alpha_2$.  Thus the Gribov horizon yields 
the Lagrangian in the nonlinear gauge $\cl_{NL}$ \cite{btm, zj, hs1}.

\section{$\alpha \neq 0$ gauge}

     In the $\alpha \neq 0$ gauge, as $\ptl_{\mu}A_{\mu}\neq 0$ and 
\[
 \int dx\, i\cb\cdot\ptl_{\mu}D_{\mu}c=\int dx\, i(D_{\mu}\ptl_{\mu}\cb)\cdot c, \quad
 \ptl_{\mu}D_{\mu}\neq D_{\mu}\ptl_{\mu},  
\]
the operator $\ptl_{\mu}D_{\mu}$ is not hermitian.  
We assume that the operator $\ptl_{\mu}D_{\mu}$ has a 
pair of zero modes $(u_0, u_0^*)$ and a real single zero mode $v_0$.  Then $c$ is expanded as 
\begin{equation}
 c(x) =  \xi u_0(x) + \xi^{\dagger}u_0^*(x) + \zeta v_0 + \cdots, \lb{301}
\end{equation}
where $\xi, \xi^{\dagger}$ and $\zeta$ are independent Grassmann variables.  
Although the Lagrangian (\ref{202}) does not contain 
$ \xi, \xi^{\dagger}$ and $\zeta$, the measure $Dc$ contains $d\xi d\xi^{\dagger} d\zeta$.  
Thus we find 
\begin{align}
    \int Dc D\cb \ e^{-\int dx \cl_{\alpha}} = 0, \notag \\
   \int Dc D\cb \ c^Ac^Bc^C\  e^{-\int dx \cl_{\alpha}} \neq 0. \lb{302}
\end{align}
However (\ref{302}) contradicts with the ghost number conservation.  To avoid 
this problem, a pair of zero modes $(\bar{u}_0, \bar{u}_0^*)$ and a real single zero mode $\bar{v}_0$ 
of the operator $D_{\mu}\ptl_{\mu}$ must exist, 
\footnote{Let us consider a square matrix $\mathcal{D}$, which is 
not necessarily hermitian.  There are eigenvectors $V_k$ which satisfy 
$\mathcal{D}V_k=\lambda_kV_k$.  Since 
$\det (\mathcal{D}-\lambda E)=\det (^t\mathcal{D}-\lambda E)$, $^t\mathcal{D}$ 
has the same eigenvalues as $\mathcal{D}$.  Thus we have 
$^t\mathcal{D}U_l=\lambda_l U_l$.  As $^tU_l$ satisfies 
$^tU_l\mathcal{D}=\lambda_l \,^tU_l$, these eigenvectors satisfy $^tU_lV_k=0$ 
if $\lambda_l\neq \lambda_k$ \cite{yk}.  In the present case, we assign 
$\mathcal{D}=\ptl_{\mu}D_{\mu}$, 
$^t\mathcal{D}=D_{\mu}\ptl_{\mu}$, $V_k=(u_k, u_k^*, v_k)$ and 
$U_l=(\bar{u}_l,\bar{u}_l^*, \bar{v}_l)$.  } 
and $\cb$ is expanded as 
\begin{equation}
 \cb(x)= \bar{\xi}\bar{u}_0 + \bar{\xi}^{\dagger}\bar{u}_0^*(x) + \bar{\zeta}\bar{v}_0(x) + \cdots. 
\lb{303}
\end{equation}
Since $\ptl_{\mu}D_{\mu}\neq D_{\mu}\ptl_{\mu}$, 
a zero-mode pair ($\bar{u}_0$, $\bar{u}_0^*$) is different from ($u_0$, $u_0^*$), and $\bar{v}_0\neq v_0$. 

     Now we consider the effect of the zero-mode pairs ($u_0$, $u_0^*$) and ($\bar{u}_0$, $\bar{u}_0^*$).  
Since the Lagrangian does not contain $\xi, \xi^{\dagger}, \bar{\xi}$ and $\bar{\xi}^{\dagger}$, 
and the measure contains $d\xi d\xi^{\dagger} d\bar{\xi} d\bar{\xi}^{\dagger}$, to obtain a 
non-zero partition function, we must repeat the consideration in \S 2.  
Namely the zero-mode pairs give rise to the effective Lagrangian $\cl_{eff}$, 
and the nonlinear gauge is realized.  

     Next we study tha terms $\zeta v_0$ in (\ref{301}) and $\bar{\zeta}\bar{v}_0$ in (\ref{303}).  
The Lagrangian $\cl_{eff}$ has the term $ig \alpha_2 B\cdot (\cb \times c)$.  
Although this term is necessary to ensure the BRS symmetry, as 
\begin{equation}
 B\cdot (\cb \times c)=B\cdot \{ \bar{\zeta}\zeta \bar{v}_0(x)\times v_0(x) + \cdots \}, \lb{304} 
\end{equation}
the partition function does not vanish even if $DcD\cb$ contains $d\zeta d\bar{\zeta}$.  

     Thus, when $\alpha \neq 0$, the partition function changes from (\ref{201}) to (\ref{218}), 
if the FP operator $\partial_{\mu}D_{\mu}$ has a pair of zero modes.  This result is unchanged 
even if this operator has a single zero mode.

\section{Renormalization group flow of $\alpha$}

     We return to the gauge $\alpha = 0$, and assume $\partial_{\mu}D_{\mu}$ has a single zero mode 
$v_0$.  Now $\partial_{\mu}D_{\mu}=D_{\mu}\partial_{\mu}$ holds, we must set $\bar{v}_0(x)=v_0(x)$ 
in (\ref{303}), i.e. 
\[
  c= \zeta v_0(x) + \cdots , \quad \cb = \bar{\zeta} v_0(x) + \cdots.  
\] 
Since $v_0(x)\times v_0(x)=0$, $\cb \times c$ and (\ref{304}) do not contain $\bar{\zeta} \zeta$.  
Namely we cannot say that $Z_{\alpha=0}^{NL} \neq 0$ is guaranteed.  

     To evade this difficulty, we first construct the partition function $Z_{\alpha}^{NL} \neq 0$,  
and then take the limit $\alpha \to 0$, i.e. $\lim_{\alpha \to 0}Z_{\alpha}^{NL}$.  

     From the Lagrangian $\cl_{NL}$, the equation of motion for $B$ is 
\[
 \ptl_{\mu} A_{\mu}-\alpha B=-ig\alpha_2 (\cb\times c). 
 \]
So, when $\alpha \to 0$, the term $-ig\alpha_2 (\cb\times c)$ 
must be taken into account.  
In this section, treating the interactions perturbatively at the one-loop level, 
we study the behavior of $\alpha$.  

     In Appendix C, we derive the renormalization group (RG) equations 
\begin{equation}
 \mu \frac{\ptl \alpha_1}{\ptl \mu} = \frac{g^2C_2(G)}{16\pi^2} \alpha_1\left(\frac{13}{3}-\alpha_1\right), \quad 
 \mu \frac{\ptl \alpha_2}{\ptl \mu} = \frac{g^2C_2(G)}{16\pi^2} \alpha_2\left(\frac{13}{3}-\alpha_2\right),  \lb{401}
\end{equation}
which coincide with the results in Refs.~\cite{hn} and \cite{kmsi}.\footnote{The parameters 
$\alpha_1$ and $\alpha_2$ in this article are related to the parameters in Refs.~\cite{hn} and \cite{kmsi} 
as follows:
\begin{enumerate}
 \item after setting $\xi=0, \zeta=\eta$ and $\alpha=\beta$, $\alpha_1=(1+\eta)\alpha$ and $\alpha_2=-\eta\alpha$ in Ref.~\cite{hn}, 
 \item $\alpha_1=(1-\xi)\lambda=\alpha'+ \alpha/2$ and $\alpha_2=\xi\lambda=\alpha/2$ in Ref.~\cite{kmsi}.
\end{enumerate}
}
We emphasize that the equation for $\alpha_1$ does not contain $\alpha_2$, and vice versa.  
From (\ref{401}),  $\alpha=\alpha_1+\alpha_2$ satisfies 
\begin{equation}
 \mu \frac{\ptl \alpha}{\ptl \mu} = \frac{g^2C_2(G)}{16\pi^2} \left\{\frac{13}{3}\alpha
-\alpha^2+2(\alpha-\alpha_2)\alpha_2\right\}.  \lb{402}
\end{equation}
When $|\alpha|\ll 1$, (\ref{402}) becomes 
\begin{equation}
 \mu \frac{\ptl \alpha}{\ptl \mu} \simeq -\frac{g^2C_2(G)}{8\pi^2} \alpha_2^2.  \lb{403}
\end{equation}
Therefore, when $\alpha_2\neq 0$, $\alpha$ increases as $\mu$ decreases.  
The quartic ghost interaction makes $\alpha\neq 0$, and the situation in \S 3 realizes.  
Even if a single zero mode $v_0$ exists, the partition function does not vanish.  

     Eq.(\ref{401}) shows that $(\alpha_1,\alpha_2)=(0,0)$ is an infrared fixed point.  Does 
this fact imply that the Landau gauge (\ref{101}) is retrieved as $\mu \to 0$?  Does 
the process in \S 2 repeat again?  In the next section, 
we show such a trouble does not happen.  

\section{Ghost condensation}

In Appendix B, we present the Lagrangian \cite{hs1,hs2}
\begin{equation}
 \cl_{\vp}=-\frac{\alpha_1}{2}B^2 
 +B(\ptl_{\mu}A_{\mu}+\vp -w)+i\cb \cdot(\ptl_{\mu}D_{\mu}+g\vp \times )c 
 +\frac{\vp^2}{2\alpha_2}.  \lb{501}
\end{equation}
This Lagrangian has the BRS invariance, if $\vp$ transforms as 
$\dl_B \vp=g\vp \times c.$  
Setting the constant $w=0$, and performing the $\vp$ integration, we find $\cl_{\vp}$ yields $\cl_{NL}$.  
Namely, $\vp$ is an auxiliary field which represents $\alpha_2 \bb$.  

     However, in a low energy region, $\vp$ is not an auxiliary field.  In Ref.~\cite{hs2}, 
we derived another RG equation for $\alpha_2$ given by 
\begin{equation}
 \mu\frac{\ptl}{\ptl \mu}\alpha_2=\frac{g^2C_2(G)}{(4\pi)^2}\left(\beta_0-2\alpha_2 \right)\alpha_2,  \lb{502}
\end{equation}
which is different from (\ref{401}).  Eq.(\ref{502}) was derived by making the Wilsonian effective action for 
$\vp$.\footnote{In Appendix C.2, we explain how to derive (\ref{502}) from $\cl_{NL}$.}  
We also showed that $\vp$ acquires the vacuum expectation value $\langle \vp\rangle =\vp_0$ 
under the energy scale 
\begin{equation}
   \mu_0=\Lambda e^{-4\pi^2/(\alpha_2 g^2)},  \lb{503}
\end{equation}
where $\Lambda$ is a momentum cut-off.  
Ghost-antighost bound states and ghost condensation 
appear below $\mu_0$.  We substitute $\vp(x)=\vp_0 + \vp'(x)$ into (\ref{501}), and choose the constant 
$w=\vp_0$.  This choice is necessary to maintain the BRS symmetry \cite{kug}.\footnote{This point is explained in Appendix D.  
The anti-BRS symmetry and the global gauge symmetry are also discussed.}  
Then (\ref{501}) becomes
\begin{equation}
 -\frac{\alpha_1}{2}B^2 
 +B(\ptl_{\mu}A_{\mu}+\vp' )+i\cb \cdot(\ptl_{\mu}D_{\mu}+g\vp' \times + g\vp_0 \times )c .
   \lb{504}
\end{equation}
Because of the dimensional transmutation \cite{gn}, the parameter below $\mu_0$ is not $\alpha_2$ but $\vp_0$.  

     Contrary to $\alpha_2$, the gauge parameter $\alpha_1$ remains in (\ref{504}).  As we explain in Appendix C.2, 
the RG equation (\ref{401}) for $\alpha_1$ persists, and $\alpha_1 = 0$ is an infrared fixed point.  
So, when $\mu \to 0$, (\ref{504}) gives the gauge condition 
\begin{equation}
   \ptl_{\mu}A_{\mu}+\vp' \approx 0  \lb{505}
\end{equation}
and the ghost Lagrangian 
\begin{align}
  \int dx \, i\cb \cdot(\ptl_{\mu}D_{\mu}+g\vp' \times )c &=  \int dx \, i\cb \cdot(D_{\mu}\ptl_{\mu} )c \notag \\
   &=  \int dx \, i (\ptl_{\mu} D_{\mu} \cb ) \cdot c.  \notag
\end{align}
As (\ref{505}) means $\ptl_{\mu} D_{\mu}\neq D_{\mu}\ptl_{\mu}$, 
we assume $\ptl_{\mu} D_{\mu}$ has a pair of zero modes $(u_0, u_0^*)$ 
and a single zero mode $v_0$, and $D_{\mu}\ptl_{\mu}$ has zero modes $(\bar{u}_0, \bar{u}_0^*)$ and $\bar{v}_0$.  
Even if the measure $Dc D\cb$ contains 
$d\xi d\xi^{\dagger}d\zeta d\bar{\xi}d\bar{\xi}^{\dagger}d\bar{\zeta}$, 
because the term $i\cb \cdot( g\vp_0 \times c)$ in (\ref{504}) has 
\begin{equation}
    -ig\vp_0 \cdot \{ \bar{\xi}\xi \bar{u}_0(x)\times u_0(x)+ 
   \bar{\xi}^{\dagger}\xi^{\dagger} \bar{u}_0^*(x)\times u_0^*(x) + \bar{\zeta}\zeta \bar{v}_0(x)\times v_0(x) + \cdots \}, 
\end{equation}
the partition function does not vanish.

\section{Summary}

     In the Landau gauge $\alpha=0$, the FP operator $-\ptl_{\mu}D_{\mu}$ 
has zero modes on the Gribov horizon.  As the ghost $c$ and the anti-ghost $\cb$ are Grassmann variables, 
it is natural to expect that these zero modes yield effective ghost interactions.  
We have shown the quartic ghost interaction is produced by a pair of zero modes.  
If we impose the BRS invariance, the Lagrangian in the nonlinear gauge is obtained.  Thus the 
Landau gauge changes to the nonlinear gauge.  
In the $\alpha \neq 0$ gauge, the same result is obtained as well.  

     The effect of a single zero mode was also studied.  Although there is no trouble in the 
$\alpha \neq 0$ gauge, the partition function $Z$ may vanish in the $\alpha = 0$ gauge.  
We can avoid this problem by taking the limit $\alpha \to 0$. 

     Usually, when $\det \ptl_{\mu}D_{\mu}=0$ for some configuration $A_{\mu}$, we can 
evade the $Z=0$ problem by choosing another gauge (locally) \cite{na}.  In this paper, 
we have shown that such a configuration changes 
the gauge to the nonlinear gauge automatically.  

     The partition functions in the Landau gauge and the nonlinear gauge are equivalent 
perturbatively.  
In the nonlinear gauge, $(\alpha_1,\alpha_2)=(0,0)$ is an infrared fixed point at the one-loop level.  
In this case, the Landau gauge is retrieved and the zero-mode problem appears again.  
However, this scenario is not true.  The nonlinear gauge yield the ghost condensation 
below the energy scale $\mu_0$, and the zero-mode problem no longer happens.

\appendix

\section{Examples of zero modes in the Coulomb gauge}

     In this appendix, choosing the gauge $\ptl_{j}A_{j}=0$, we study the eigenvalue equation 
\begin{equation}
    -\ptl_{j}D_{j}u=-(\triangle +gA_{j}\times \ptl_j)u= \lambda u  \lb{a01}
\end{equation}
in three-dimensional space-time.  

\subsection{A pair of zero modes}

     If the eigenfunction has the form $u^A=e^{is}w^A$ with $gA_{j}\times (\ptl_j w)=0$, 
(\ref{a01}) becomes 
\begin{equation}
    -iH^{AB}e^{is}w^B = (\triangle + \lambda )e^{is}w^A ,\quad 
    H^{AB} = gf^{ACB}A_j^C (\ptl_j s) .  \lb{a02}
\end{equation}
Since $H$ is a real antisymmetric $3\times 3$ matrix, its eigenvalues are pure imaginary or $0$, i.e. 
\begin{equation}
     H^{AB}w_+^B = ih(x)w_+^A,\  H^{AB}w_-^B = -ih(x)w_-^A,\ H^{AB}w_0^B = 0.  \lb{a03}
\end{equation}
The last equation of (\ref{a03}) means that the effect of $A_j^C$ disappears and $w_0$ does not 
become a zero mode.  
From (\ref{a02}) and (\ref{a03}), we obtain 
\[  h(x)e^{\pm is}w_{\pm}^A = (\triangle + \lambda )e^{\pm is}w_{\pm}^A.   \]
Thus we find the two functions $u_{\pm}=e^{\pm is}w_{\pm}$ become a zero-mode pair, if 
\begin{equation}
   h(x)u_{\pm}^A = \triangle u_{\pm}^A   \lb{a04}
\end{equation}
holds.  

     To give concrete examples, let us choose the abelian configuration 
\begin{equation}
  A_i^A(\vec{x}) = a_i(\vec{x})\delta^{A3},  \quad \ptl_ia_i=0.  \lb{a05}
\end{equation}

\subsubsection{Three-torus $T^3$}

    Gribov copies in the three-torus $T^3$ are studied in Ref.~\cite{vb}.  The constant configuration 
\[
  a_j(x) =  \frac{C_j}{gL}, \quad C_1=2\pi,\ -2\pi<C_2<2\pi,\ -2\pi<C_3<2\pi,  
\]
is on the first Gribov horizon, where $L$ is the size of the torus.  Setting $s=2\pi x_1/L$, we find 
(\ref{a04}) is satisfied by a zero-mode pair 
\[ u_{\pm}=e^{\pm i2\pi x_1/L} \begin{pmatrix}1\cr \pm i \cr 0 \cr \end{pmatrix}.  \]

\subsubsection{Axially symmetric configuration in $\mathbb{R}^3$}

     Next we consider the configuration 
\begin{equation}
  a_j(x)=\epsilon_{j3k}q(r)x_k,  \lb{a06}
\end{equation}
where $(r, \theta, \phi)$ are the spherical coordinates.  
Using the angular momentum operator $\hat{L}_j=-i\epsilon_{jkl} x_k \ptl_l$, we 
find 
\[ -\triangle =-\frac{1}{r^2}\frac{\ptl}{\ptl r}r^2\frac{\ptl}{\ptl r}
+\frac{\hat{L}^2}{r^2}, \quad H^{AB} = gf^{A3B} q(r) i\hat{L}_3 s.  \]
Then it is natural to set $e^{is}=e^{im\phi}$ and 
\[ w_+=i^lR(r)\Theta_{lm}(\theta)\begin{pmatrix}1\cr i \cr 0 \cr \end{pmatrix}, \]
where $l$ and $m$ are integers, and 
\[ \Theta_{lm}(\theta)= \frac{(-1)^m }{\sqrt{2\pi}}\sqrt{\frac{(2l+1)(l-m)!}{2(l+m)!}}P_l^m(\cos \theta). \] 
We note $e^{im\phi}\Theta_{lm}(\theta)=Y_{lm}(\theta,\phi)$ is the spherical harmonics which satisfies 
$Y_{lm}^*=(-1)^{-m}Y_{l,-m}$.  Then (\ref{a04}) becomes 
\begin{equation}
 \left[ -\frac{1}{r^2}\frac{\ptl}{\ptl r}r^2\frac{\ptl R(r)}{\ptl r}+
\frac{l(l+1)}{r^2}R(r)- gm q(r) R(r) \right] Y_{l,\pm m}(\theta, \phi)=0.  
 \lb{a07}
\end{equation} 

     Now, following Henyey \cite{he}, we substitute the functions 
\begin{equation}
 R(r)=\frac{K r^{\rho}}{(r^2+r_0^2)^{\kappa}}, \quad 
q(r)=\frac{d}{(r^2+r_0^2)^{\sigma}}  \lb{a08}
\end{equation}
into (\ref{a07}), where $K, r_0, d, \rho, \kappa$ and $\sigma$ are constants.  
Eq.(\ref{a07}) is satisfied by 
\[ \sigma=2,\ \rho=l,\ \kappa=l+\frac{1}{2},\ d=\frac{(2l+1)(2l+3)}{gm}r_0^2.  \]
Thus we obtain the abelian configuration and the corresponding zero-mode pairs as 
\begin{align}
  a_j&= \frac{(2l+1)(2l+3)}{gm} \frac{r_0^2}{(r^2+r_0^2)^2}\epsilon_{j3k}x_k, 
\notag \\
u_{\pm}&= i^l\frac{Kr^l}{(r^2+r_0^2)^{l+1/2}}Y_{l,\pm m}(\theta,\phi)  
\begin{pmatrix}1 \cr \pm i \cr 0 \cr \end{pmatrix}.  \quad (l\geq 1, m=1,2,\cdots,l)  
\lb{a10}
\end{align}
In Ref.~\cite{he}, the $l=1$ case is presented explicitly.

\subsection{A single zero mode}

     In Ref.~\cite{ma}, a single zero mode was found in an instanton background.  Here we 
give an example in $\mathbb{R}^3$.  Generalizing (\ref{a05}) and (\ref{a06}), 
we choose the configuration 
\begin{equation}
     A_j^C(x)=\epsilon_{jCk}q(r)x_k.  \lb{a11}  
\end{equation}
Then (\ref{a01}) becomes 
\begin{align}
  -igq(r)\Xi^{AB}u^B &= (\triangle + \lambda) u^A,  \lb{a12} \\
  gf^{ACB}A_j^C\ptl_j &=igq(r)\Xi^{AB},\quad \Xi^{AB}=f^{ACB}\hat{L}^C. \notag 
\end{align}

     First we solve the equation 
\begin{equation}
     \Xi^{AB} u^B=i\alpha u^A,  \lb{a13} 
\end{equation}
where $i\alpha$ is an eigenvalue of $\Xi$.  We substitute the expansion 
\[ u^A=\sum_{m=-l}^lR^A_{lm}(r)Y_{lm}(\theta,\phi), \]
and, for simplicity, choose $l=1$.  
Then we find that the eigenvalues are $\alpha=2, 1$ and $-1$, 
and the numbers of eigenfunctions are $1, 3$ and $5$, respectively.  
We choose the real eigenfunctions $u_{\alpha}^A=R_{\alpha}(r)w_{\alpha}^A(\theta,\phi)$, where 
$w_{\alpha}^A(\theta,\phi)$ are given by 
\begin{align*}
 \alpha=2 &:\ \begin{pmatrix}Y_{11}-Y_{1,-1}\cr -i(Y_{11}+Y_{1,-1}) \cr -\sqrt{2}Y_{10} \cr \end{pmatrix},  \\
 \alpha=1 &:\ \begin{pmatrix}\sqrt{2}Y_{10} \cr 0 \cr Y_{11}-Y_{1,-1}\cr \end{pmatrix},
 \ \begin{pmatrix} 0 \cr \sqrt{2}Y_{10} \cr -i(Y_{11}+Y_{1,-1}) \cr \end{pmatrix},
 \ \begin{pmatrix} i(Y_{11}+Y_{1,-1})\cr Y_{11}-Y_{1,-1} \cr 0 \cr \end{pmatrix},  \\
 \alpha=-1 &:\  \begin{pmatrix}\sqrt{2}Y_{10} \cr 0 \cr -(Y_{11}-Y_{1,-1})\cr \end{pmatrix},
 \ \begin{pmatrix} 0 \cr \sqrt{2}Y_{10} \cr i(Y_{11}+Y_{1,-1}) \cr \end{pmatrix},\\
   &\ \ \ \begin{pmatrix} Y_{11}-Y_{1,-1}\cr i(Y_{11}+Y_{1,-1}) \cr 0 \cr \end{pmatrix},
 \ \begin{pmatrix} i(Y_{11}+Y_{1,-1})\cr -(Y_{11}-Y_{1,-1}) \cr 0 \cr \end{pmatrix},
 \ \begin{pmatrix} Y_{11}-Y_{1,-1} \cr -i(Y_{11}+Y_{1,-1}) \cr 2\sqrt{2}Y_{10} \cr \end{pmatrix}.  
\end{align*}

     Next we determine $R_{\alpha}$.  From (\ref{a12}) with $\lambda=0$ and (\ref{a13}), $R_{\alpha}$ satisfies  
\begin{equation}
  -\frac{1}{r^2}\frac{\ptl}{\ptl r}r^2\frac{\ptl R_{\alpha}(r)}{\ptl r}+
\frac{l(l+1)}{r^2}R_{\alpha}(r)+ g\alpha q(r) R_{\alpha}(r) =0.  
 \lb{a14}
\end{equation} 
As in the previous subsection, we substitute (\ref{a08}) into (\ref{a14}).  Then we find 
\begin{equation}
 R_{\alpha}(r)=R(r)=\frac{K r}{(r^2+r_0^2)^{3/2}}, \quad 
q(r)=\frac{-15}{\alpha g}\frac{r_0^2}{(r^2+r_0^2)^2}.  \lb{a15}
\end{equation}
     
     Two real zero modes are replaced by a pair of zero modes.  So one real zero mode remains 
for each value of $\alpha$.

\section{Derivation of the Lagrangians (\ref{219}) and (\ref{501}) by the use of "source"}

     In the instanton case, the fermion determinant does not vanish if fermion sources exist \cite{tH,tH2}.  
Following this case, we introduce a field $\vp(x)$, and replace $i\cb\cdot \ptl_{\mu}D_{\mu}c$ with 
\begin{equation}
 i\cb\cdot[ \ptl_{\mu}D_{\mu}+g \vp \times ]c. \lb{e1}
\end{equation}

The eigenvalue equation is 
\[ -[ \ptl_{\mu}D_{\mu}+g \vp \times ]w_n= \Lambda_n w_n. \]
We treat the term $g \vp \times $ as pertubation, and perform 
the expansion 
\[ w_n=w_n^{(0)}+w_n^{(1)}+\cdots, \quad \Lambda_n =\Lambda_n^{(0)}+
\Lambda_n^{(1)}+\cdots,  \]
where $\Lambda_n^{(0)}=\lambda_n$, and $w_n^{(0)}=u_n$ in (\ref{204}) and $w_n^{(0)}=u_n^*$ in (\ref{205}).  
Using the normalization 
$\int dx\, u_n^*\cdot u_n=1$ and $\int dx\, u_n \cdot u_n=0$, we obtain 
\[ \Lambda_n^{(1)}=g \int dx\, u_n^*\cdot (\vp \times u_n)  \]
for $u_n$ and 
\[ \Lambda_n^{(1)*}=g \int dx\, u_n \cdot (\vp \times u_n^*)  \]
for $u_n^*$, where $f^{ABC} u_n^A \vp^B u_n^C=0$ has been used.  
Therefore, if $\ptl_{\mu}D_{\mu}$ has a pair of zero modes $(u_0, u_0^*)$, 
(\ref{e1}) gives rise to the determinant 
\begin{eqnarray}
 \det [-\ptl_{\mu}D_{\mu}-g \vp \times ]=\prod_{n}|\Lambda_n|^{k_n}
 &\approx& |\Lambda_0^{(1)}|^2 \prod_{n\neq 0} |\Lambda_n|^{k_n} \nonumber \\
 &=& \left|g \int dx\, u_0^*\cdot(\vp\times u_0)\right|^2
\prod_{n\neq 0} |\Lambda_n|^{k_n},  \lb{e2}
\end{eqnarray}
where $k_n$ is the number of eigenfunctions that have the eigenvalue $\Lambda_n$ or $\Lambda_n^*$.  
Thus, although $\Lambda_0^{(0)}=\lambda_0=0$, $\Lambda_0^{(1)}\neq 0$ makes 
the partition function non-zero.  

     Since 
\begin{eqnarray*}
\left|g \int dx\, u_0^*\cdot(\vp\times u_0)\right|^2 &\propto & 
\int d\xi d\bar{\xi} d\xi^{\dagger} d\bar{\xi}^{\dagger} 
\left[g \int dx\, \bar{\xi}^{\dagger}u_0^*\cdot(\vp\times \xi u_0) \right] \cr
& &\times \left[g \int dy\, \bar{\xi}u_0\cdot(\vp\times \xi^{\dagger} u_0^*)\right], 
\end{eqnarray*}
we find 
\begin{equation}
  DcD\cb \exp\left\{-i\int dx\, \cb \cdot(\ptl_{\mu}D_{\mu}+g \vp
\times ) c\right\}     \lb{e3}
\end{equation}
gives the determinant (\ref{e2}).  
To derive (\ref{219}), we multiply (\ref{e3}) by $\exp[-\int dx\, (\vp+\alpha_2 B)^2/(2\alpha_2)]$, and integrate 
with respect to $\vp$:  
\[
 D\varphi\,\exp\left\{-i\int dx\,\cb \cdot[\ptl_{\mu}D_{\mu}+g\vp\times] c
-\int dx
\, \left(\frac{\vp^2}{2\alpha_2}+B\vp +\frac{\alpha_2}{2}B^2\right)\right\}.  
\]
After the $\vp$ integration, we obtain (\ref{219}).  

     We note, to derive (\ref{501}), (\ref{e3}) must be multiplied by 
$\exp[-\int dx\, \left\{(\vp-w+\alpha_2 B)^2+2w\vp-w^2\right\}/(2\alpha_2)]$, where $w$ is a constant determined later.

\section{Derivation of the RG equations (\ref{401}) and (\ref{502})}

     In subsection C.1,  using $\cl_{NL}$, we derive the RG equation (\ref{401}).  In 
subsection C.2, the RG equation (\ref{502}) is derived.  The RG equation for $\alpha_1$ 
under the scale $\mu_0$ is discussed.  

\subsection{The Lagrangian (\ref{219}) and the RG equations (\ref{401})}    

\subsubsection{Equation for $\alpha_2$}

     The Lagrangian $\cl_{NL}$ contains the quartic ghost interaction 
\[ -\frac{\alpha_2}{2}(ig \cb \times c)^2.   \]
We define the renormalization constant $Z_4$ by 
\begin{equation}
     (\alpha_2 g^2)_0 =Z_4 \tilde{Z}_3^{-2} \alpha_2 g^2,  \lb{c01}
\end{equation}
where $\cb_0=\tilde{Z}_3^{1/2}\cb$ and $c_0=\tilde{Z}_3^{1/2}c$.  First we consider the 
ghost self-energy.  
Although $\cl_{NL}$ gives additional one-loop diagrams, divergence of them cancels out.  
Thus we obtain, as usual, $\tilde{Z_3}=1+\tilde{Z_3}^{(1)}+\cdots$ with 
\begin{equation}
     \tilde{Z_3}^{(1)}=\frac{2g^2}{(4\pi)^2}\left(3-\alpha \right)\frac{1}{4\varepsilon}, \lb{c02}
\end{equation}
where $\varepsilon=(4-D)/2$, and $C_2(G)=2$ is inserted.  We note the gauge parameter in $\cl_{NL}$ 
is $\alpha=\alpha_1 +\alpha_2$.  

\begin{figure}
\begin{center}
\includegraphics{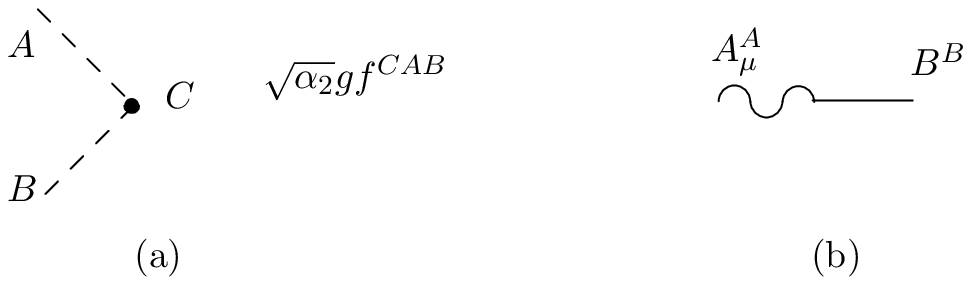}
\caption{The vertex and the propagator peculiar to $\cl_{NL}$.}
\label{fig1}
\end{center}
\end{figure}

     Next we study $Z_4$.  Using the notation of Fig.C1, one-loop diagrams which contribute to 
$Z_4$ come from the diagrams in Figs.C2 and C3.  However Fig.C2(b) does not yield divergence, 
and divergences of Figs.C2(c1)-(c3) cancel out.  Furthermore some of the diagrams derived from Fig.C3 don't diverge.  

\begin{figure}
\begin{center}
\includegraphics{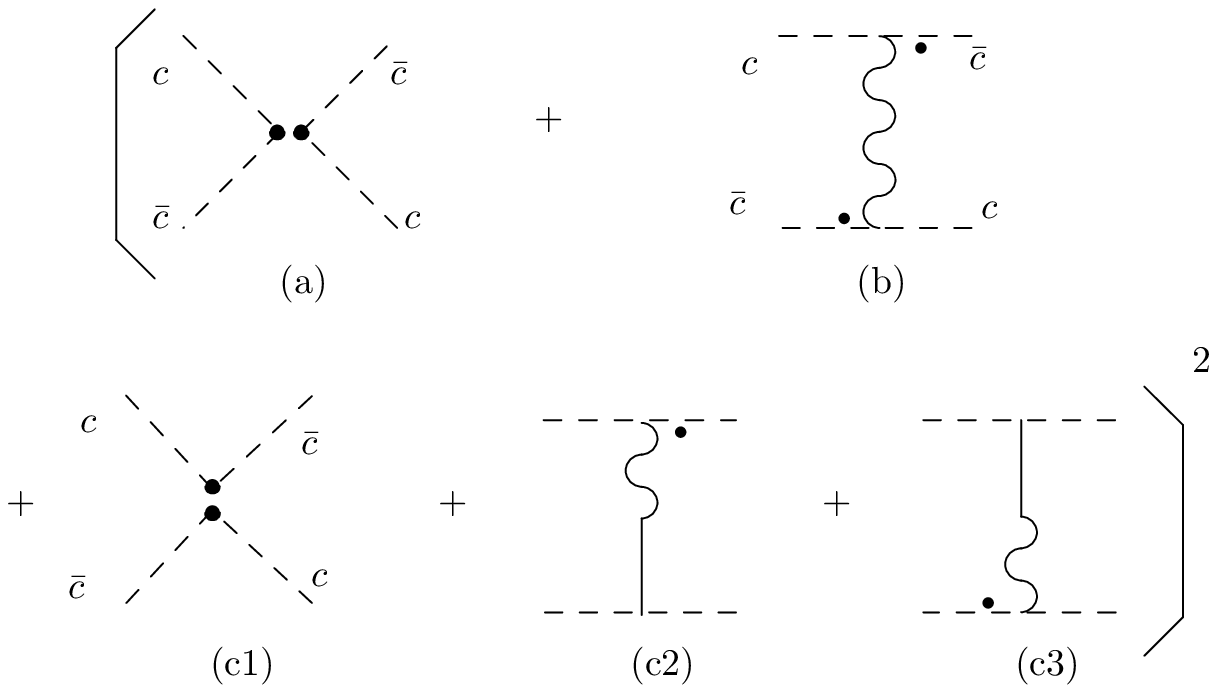}
\caption{The diagrams which contribute to one-loop correction for $(\cb \times c)^2$ vertex.}
\label{fig2}
\end{center}
\end{figure}

\begin{figure}
\begin{center}
\includegraphics{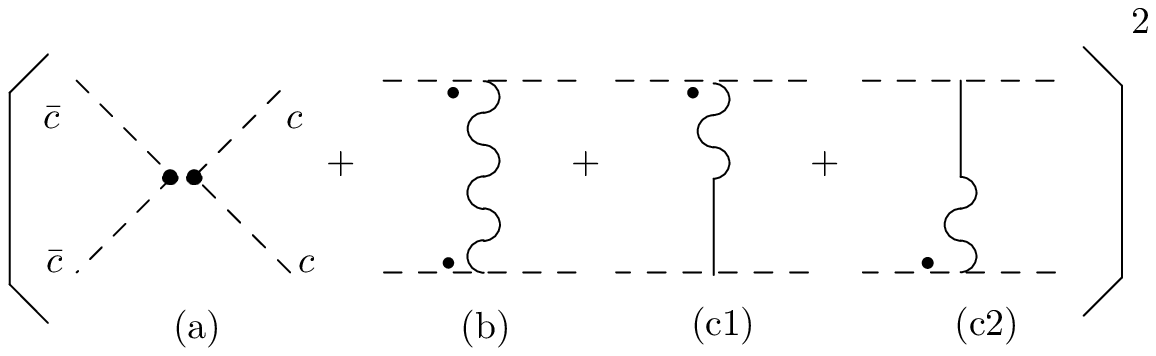}
\caption{The diagrams which contribute to one-loop correction for $(c\times c)(\cb \times \cb)$ vertex.}
\label{fig3}
\end{center}
\end{figure}

Thus divergent diagrams are depicted in Fig.C4, and they give the constant 
$Z_4=1+Z_{4a}^{(1)}+Z_{4b}^{(1)}+Z_{4c}^{(1)}+Z_{4d}^{(1)}+\cdots$, where 
\begin{align}
 Z_{4a}^{(1)}&= \frac{2g^2}{(4\pi)^2}\left(-\alpha_2\right)\frac{1}{\varepsilon}, & 
 Z_{4b}^{(1)}&=\frac{2g^2}{(4\pi)^2}\left(-\alpha_2\right)\frac{1}{2\varepsilon},  \notag \\
 Z_{4c}^{(1)}&=\frac{2g^2}{(4\pi)^2}\left(\alpha_2\right)\frac{1}{\varepsilon},& 
 Z_{4d}^{(1)}&=\frac{2g^2}{(4\pi)^2}\left(-\alpha_1-\alpha_2\right)\frac{1}{2\varepsilon}. \lb{c03} 
\end{align}
Eq.(\ref{c01}) leads to 
\begin{equation}
\mu\frac{\ptl \alpha_2 g^2}{\ptl \mu}=-\frac{\mu}{Z_4\tilde{Z}_3^{-2}}\frac{\ptl Z_4\tilde{Z}_3^{-2}}{\ptl \mu}\alpha_2g^2, 
\ Z_4\tilde{Z}_3^{-2}=1+Z_{4a}^{(1)}+Z_{4b}^{(1)}+Z_{4c}^{(1)}+Z_{4d}^{(1)}-2\tilde{Z_3}^{(1)}+\cdots.  \lb{c04}
\end{equation}
Then performing the replacement $g \to g\mu^{-\varepsilon}$ or $1/\varepsilon \to 2\ln \Lambda/\mu$ in (\ref{c02}) and (\ref{c03}), 
and using the RG equation 
\[  \mu\frac{\ptl g}{\ptl \mu}=-\beta_0\frac{g^3}{(4\pi)^2}, \quad \beta_0=\frac{22}{3}, \]
we obtain 
\begin{equation}
   \mu\frac{\ptl}{\ptl \mu}\alpha_2=\frac{2g^2}{(4\pi)^2}\left(\frac{13}{3}-\alpha_2 \right)\alpha_2.  \lb{c05}
\end{equation}

\begin{figure}
\begin{center}
\includegraphics{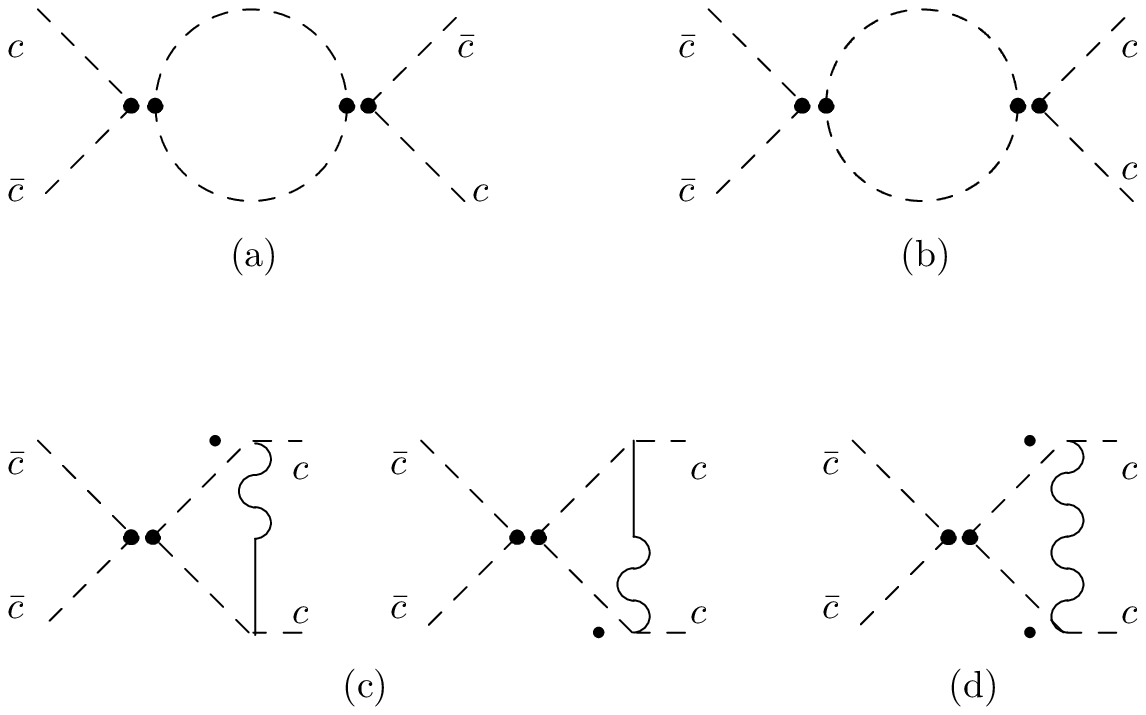}
\caption{The one-loop divergent diagrams for $(\cb \times c)^2$ vertex.}
\label{fig4}
\end{center}
\end{figure}

\subsubsection{Equation for $\alpha_1$}

     Renormalization constants are defined as usual:
\begin{align}
 &A_0^{\mu}=\sqrt{Z_3}A^{\mu},\ Z_3 =1+Z_3^{(1)}+\cdots,\quad B_0=\sqrt{Z_B}B,\ Z_B =1+Z_B^{(1)}+\cdots, \notag \\
 &(\alpha_j)_0=Z_{\alpha_{j}}\alpha_j,\ Z_{\alpha_{j}} =1+Z_{\alpha_{j}}^{(1)}+\cdots, (j=1,2).  \lb{c06}
\end{align}
Then $\cl_{NL}$ gives the counter terms 
\[ \frac{1}{2}(Z_B^{(1)}+Z_3^{(1)})B \ptl_{\mu}A^{\mu},  \quad
\frac{1}{2}\left\{(Z_B^{(1)}+Z_{\alpha_{1}}^{(1)})\alpha_1 +(Z_B^{(1)}+Z_{\alpha_{2}}^{(1)})\alpha_2 \right\}B^2.  \]
The first counter term cancels the divergence of Fig.C5(a), and we obtain 
\[ Z_B^{(1)}+Z_3^{(1)}= \frac{2g^2}{(4\pi)^2}\frac{-\alpha_2}{\varepsilon}. \]
As the gauge parameter in $\cl_{NL}$ is $\alpha$, the constant $Z_3^{(1)}$ is 
\begin{equation}
  Z_3^{(1)}= \frac{2g^2}{(4\pi)^2}\left(\frac{13}{3}-\alpha \right)\frac{1}{2\varepsilon}  \lb{c07}
\end{equation}
as usual.  
Using these results, $Z_B^{(1)}$ becomes 
\begin{equation}
    Z_B^{(1)}= \frac{2g^2}{(4\pi)^2}\left(\alpha_1 - \alpha_2-\frac{13}{3} \right)\frac{1}{2\varepsilon}. \lb{c08}
\end{equation}

\begin{figure}
\begin{center}
\includegraphics{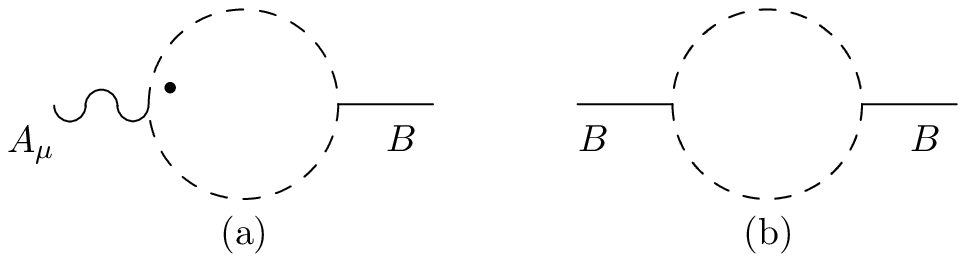}
\caption{The one-loop diagrams which contribute to the propagators for $A_{\mu}^AB^B$ and $B^AB^B$.}
\label{fig5}
\end{center}
\end{figure}

     The divergence of Fig.C5(b) is canceled by the second counter term, i.e. 
\begin{equation}
  Z_B^{(1)}(\alpha_1+\alpha_2)+Z_{\alpha_1}^{(1)}\alpha_1 +Z_{\alpha_2}^{(1)}\alpha_2= 
\frac{2g^2}{(4\pi)^2}\frac{-\alpha_2^2}{\varepsilon}.  \lb{c09}
\end{equation}
From (\ref{c05}) and (\ref{c06}), 
\[  \mu\frac{\ptl}{\ptl \mu}\alpha_2=-\mu\frac{\ptl Z_{\alpha_{2}}^{(1)}}{\ptl \mu}\alpha_2 =
\frac{2g^2}{(4\pi)^2}\left(\frac{13}{3}-\alpha_2 \right)\alpha_2 \]
and 
\begin{equation}
 Z_{\alpha_{2}}^{(1)}\alpha_2=\frac{2g^2}{(4\pi)^2}\left(\frac{13}{3}-\alpha_2 \right)\frac{\alpha_2}{2\varepsilon}  \lb{c10}
\end{equation}
is derived.  Substituting (\ref{c08}) and (\ref{c10}) into (\ref{c09}), we obtain 
\[
 Z_{\alpha_{1}}^{(1)}\alpha_1=\frac{2g^2}{(4\pi)^2}\left(\frac{13}{3}-\alpha_1 \right)\frac{\alpha_1}{2\varepsilon}  
\]
and 
\begin{equation}
   \mu\frac{\ptl}{\ptl \mu}\alpha_1=\frac{2g^2}{(4\pi)^2}\left(\frac{13}{3}-\alpha_1 \right)\alpha_1.  \lb{c11}
\end{equation}

\subsection{RG equations near $\mu_0$ and under $\mu_0$}

\subsubsection{Eq.(\ref{502})}

     The RG equation (\ref{502}) is derived from the Lagrangian $\cl_{\varphi}$ \cite{hs2}.  To derive it 
from the Lagrangian $\cl_{NL}$, we must replace (\ref{c04}) with 
\begin{equation}
     Z_4\tilde{Z}_3^{-2}\approx 1+Z_{4a}^{(1)}.  \lb{c12}
\end{equation}
Namely, in the region $\mu_0 < \mu < \Lambda$, the interaction between $\cb$ and $c$ becomes strong, and 
Fig.C4(a) is the main contribution.  In the limit $\mu \to \mu_0$, $\cb$ and $c$ make bound states and ghost condensate.  

\subsubsection{RG equation for $\alpha_1$}

     Near $\mu_0$, as we stated above, the Lagrangian (\ref{501}) should be used.  
Under $\mu_0$, we must use the Lagrangian (\ref{504}).  In these Lagrangians, 
the gauge parameter for $A_{\mu}$ is not $\alpha$ but $\alpha_1$.  
Then, instead of (\ref{c07}), we must use 
\[ Z_3^{(1)}=\frac{2g^2}{(4\pi)^2}\left(\frac{13}{3}-\alpha_1 \right)\frac{1}{2\varepsilon}. \]
Since the self-energies for $BA_{\mu}$ and $BB$ don't have divergence now, $Z_{\alpha_1}=Z_B^{-1}=Z_3$ holds.  
Thus we have 
\[ \mu\frac{\ptl}{\ptl \mu}\alpha_1=-\frac{\mu}{Z_3}\frac{\ptl Z_3}{\ptl \mu}\alpha_1=
\frac{2g^2}{(4\pi)^2}\left(\frac{13}{3}-\alpha_1 \right)\alpha_1.  \]
That is,  the RG equation for $\alpha_1$ is unchanged.

\section{Symmetries of the Lagrangian $\cl_{\varphi}$ in (\ref{501})}

\subsection{BRS symmetry}

     It is easy to check that $\cl_{\vp}$ is invariant under the BRS transformation 
\[ \dl_B A_{\mu}=D_{\mu}c,\  \dl_B c=-\frac{g}{2} c\times c, \ 
 \dl_B \cb =iB, \ \dl_B B=0, \ \dl_B \vp=g\vp \times c. \]
The constant $w$ is determined to conserve this symmetry.  From the partition function 
\[ Z_{\vp}= \int D\mu e^{-\int dx(\cl_{inv}+\cl_{\vp})}, \]
we can derive the equation of motion for $B$ as 
\begin{equation}
  \langle (-\alpha_1 B + \ptl_{\mu}A_{\mu} + \vp -w)\rangle=0, \lb{d01}
\end{equation}
where 
\[ \langle \Phi \rangle = \frac{1}{Z_{\vp}}\int D\mu \Phi e^{-\int dx(\cl_{inv}+\cl_{\vp})}. \]
Since $D\mu$ and $\cl_{inv}+\cl_{\vp}$ are invariant under the BRS transformation, 
\begin{equation}
  \langle \dl_B \Phi \rangle =0 \lb{d02}
\end{equation}
holds.  We substitute $B=-i\dl_B \cb$ and $\vp(x)=\vp_0+\vp'(x)$ into (\ref{d01}),   
and use $\langle A_{\mu}\rangle =0$, $\langle \vp'\rangle =0$.  Then (\ref{d01}) leads to 
$i\alpha_1 \langle \dl_B \cb\rangle =w-\vp_0$.  The consistency with (\ref{d02}) requires $w=\vp_0$.

\subsection{Anti-BRS symmetry}

     The anti-BRS transformation is given by 
\[ \dlb_B  A_{\mu}=D_{\mu}\cb, \ \dlb_B \cb=-\frac{g}{2} \cb\times \cb, \ 
 \dlb_B c =i\bb, \ \dlb B=gB\times \cb, \ \dlb_B \vp=0.  \]
When $\vp_0\neq 0$, from the 
equation of motion for $\vp$, $\langle \alpha_2 \bb \rangle=\langle \vp \rangle \neq 0$ 
holds.  Therefore the anti-BRS symmetry is broken spontaneously, because 
\[ \langle \dlb_B c \rangle =\langle i\bb \rangle \neq 0.  \]
In addition, we must set $w=\vp_0\neq 0$ to maintain the BRS symmetry.  As 
$\dlb_B \cl_{\vp}=-g(B\times \cb)\cdot w$, 
the Lagrangian does not respect the anti-BRS symmetry.  

\subsection{Global gauge symmetry}

     Using the constant small parameter $\theta$, the global gauge transformation is defined by 
$\dl_{\theta} \Phi = \theta \times \Phi$, where $\Phi$ represents all the fields in $\cl_{\vp}$.  
This symmetry breaks down just like the anti-BRS symmetry.  
In fact, $\vp_0\neq 0$ gives $\langle \dl_{\theta}\vp\rangle =\theta\times \vp_0$. 
and $w=\vp_0$ brings $\dl_{\theta} \cl_{\vp}=-w\cdot ( \theta \times B)$.  

    Next we study the partition function $Z_{\vp}$.  It transforms as 
$\dl_{\theta}Z_{\vp}\propto \langle \dl_{\theta}\cl_{\vp} \rangle$.  
Using $B=-i\dl_B\cb$ and (\ref{d02}), we find 
\[ \dl_{\theta}Z_{\vp}\propto -i (w \times\theta)\cdot \langle \dl_B \cb \rangle=0.  \]
Namely, because of the BRS symmetry, $Z_{\vp}$ remains invariant under this symmetry.  

     In the same way, we can show that the breaking by $w$ cannot be observed in 
any function $\langle \Psi(\Phi) \rangle$, if $\Psi(\Phi)$ is BRS-invariant.  
To show this, we consider the function 
\begin{equation}
     \langle  \dl_{\theta}\cl_{\vp} \Psi(\Phi) \rangle,  \lb{d03}
\end{equation}
which appears in $\dl_{\theta}\langle \Psi(\Phi) \rangle$.  Using $\dl_{\theta}\cl_{\vp}=-i(w\times \theta)\cdot \dl_B \cb$ 
and $\dl_B\Psi(\Phi)=0$, we find (\ref{d03}) vanishes.  Thus BRS invariant Green functions aren't broken by $\dl_{\theta}\cl_{\vp}$.


\begin{thebibliography}{99}
 \bibitem{gri}
     V.~N.~Gribov, Nucl.\ Phys.\ B {\bf 139}, 1 (1978).
 \bibitem{masnak}
     T.~Maskawa and H.~Nakajima, Prog.\ Theor.\ Phys. {\bf 60}, 1526 (1978).
 \bibitem{bal}
     P.~van Baal, Nucl.\ Phys.\ B\ \textbf{369} , 259 (1992).
 \bibitem{zw}
     D.~Zwanziger, Nucl.\ Phys.\ B\ \textbf{323} , 513 (1989).
 \bibitem{vb}
     P.~van Baal, hep-th/9511119.
 \bibitem{fuj}
     K.~Fujikawa, Prog.\ Theor.\ Phys.\ \textbf{61} , 627 (1979).
 \bibitem{hir}
     P.~Hirschfeld, Nucl.\ Phys.\ B\ \textbf{157} , 37 (1979).
 \bibitem{flpr}
     R.~Friedberg, T.~D.~Lee, Y.~Pang and H.~C.~Ren, Ann.\ Phys.\ 
     \textbf{246} , 381 (1996).
 \bibitem{fuj2}
     K.~Fujikawa, Nucl.\ Phys.\ B\ \textbf{462} , 437 (1996).
 \bibitem{ms}
     N.~Maggiore and M.~Schaden, Phys.\ Rev.\ D\ \textbf{50} , 6616 (1994).
 \bibitem{zw2}
     D.~Zwanziger, Prog.\ Theor.\ Phys.\ Suppl.\ \textbf{131} , 233 (1998).
 \bibitem{fuj3}
     K.~Fujikawa, Nucl.\ Phys.\ B\ \textbf{223} , 218 (1983).
 \bibitem{tH}
     G.~'t Hooft, Phys.\ Rev.\ Lett.\ \textbf{37} , 8 (1976).
 \bibitem{tH2}
     G.~'t Hooft, Phys.\ Rev.\ D\ \textbf{14}, 3432 (1976).
 \bibitem{cr}
     R.~J.~Crewther, Acta Physica Austriaca, Suppl. XIX , 47 (1978). 
 \bibitem{btm}
     L.~Baulieu and J.~Thierry-Mieg, Nucl.\ Phys.\ B\ \textbf{197} , 477 (1982).
 \bibitem{zj}
     J.~Zinn-Justin, Nucl.\ Phys.\ B\ \textbf{246} , 246 (1984).
 \bibitem{hs1}
     H.~Sawayanagi, Phys.\ Rev.\ D\ \textbf{67} , 045002 (2003).
 \bibitem{yk}
      K.~Yoshida and T.~Katoh, \textit{Applied Mathematics I} (Shokabo Publishing Co., Tokyo, 1961)  (in Japanese).
 \bibitem{hn}
     H.~Hata and I.~Niigata, Nucl.\ Phys.\ B\ \textbf{389}, 133 (1993).
 \bibitem{kmsi}
     K.~-I.~Kondo, T.~Murakami, T.~Shinohara and T.~Imai,  Phys.\ Rev.\ D\ \textbf{65} , 085034 (2002).
 \bibitem{hs2}
     H.~Sawayanagi,  Prog.\ Theor.\ Phys.\ \textbf{117}, 305 (2007). 
 \bibitem{kug}
     T.~Kugo, \textit{Quantum Theory of Gauge Fields}, (Baifukan, Tokyo, 1989) (in Japanese).
 \bibitem{gn}
     D.~J.~Gross and A.~Neveu, Phys.\ Rev.\ D\ \textbf{10}, 3235 (1974).
 \bibitem{na}
     V.~P.~Nair, \textit{Quantum Field Theory}, (Springer, 2005). 
 \bibitem{he}
     F.~S.~Henyey, Phys.\ Rev.\ D\ \textbf{20} (1979), 1460.
 \bibitem{ma}
     A.~Maas, Eur.\ Phys.\ J.\ C\ \textbf{48}, 179 (2006).
\end{thebibliography}
\end{document}